\documentclass[intlimits,twoside,a4paper]{article}

\usepackage{amsmath,amssymb}
\usepackage{graphicx}
\usepackage{wrapfig}
\usepackage{color}

\usepackage[T2A]{fontenc}
\usepackage[cp1251]{inputenc}

\usepackage[eqsecnum]{cmpj2}

\issue{2013}{16}{3}{33601}
\doinumber{10.5488/CMP.16.33601}

\title[Effect of polymer concentration and length of end block]%
{Effect of polymer concentration and length of
hydrophobic end block on the unimer-micelle transition broadness in
amphiphilic ABA symmetric triblock copolymer
solutions%
}
\author[X.-G. Han, Y.-H. Ma, S.-L. Ouyang]{X.-G. Han\refaddr{label1,label2}\footnote{E-mail: xghan0@163.com, Phone:
011+086-472-5954303, Fax: 011+086-472-5954303}\ , Y.-H.
Ma\refaddr{label1,label2}, S.-L. Ouyang\refaddr{label2}}

\addresses{
\addr{label1} School of Mathematics, Physics and Biological
engineering, Inmongolia Science and Technology University, Baotou
014010, China
\addr{label2} Key laboratory of Integrated
Exploitation of Bayan Obo Multi-Metal Resources, Inmongolia Science
and Technology University, Baotou 014010, China }

\date{Received December 7, 2012, in final form March 28, 2013}
\authorcopyright{X.-G. Han, Y.-H. Ma, S.-L. Ouyang, 2013}

\begin{document}

\maketitle

\begin{abstract}
The effects of the length of each hydrophobic end block $N_\textrm{st}$ and
polymer concentration $\bar{\phi}_{\mathrm{P}}$ on the transition
broadness in amphiphilic ABA symmetric triblock copolymer solutions
are studied using the self-consistent field lattice model. When the
system is cooled, micelles are observed, i.e.,the homogenous
solution (unimer)-micelle transition occurs. When $N_\textrm{st}$ is
increased, at fixed $\bar{\phi}_{\mathrm{P}}$, micelles occur at higher
temperature, and the temperature-dependent range of micellar
aggregation and half-width of specific heat peak for unimer-micelle
transition increase monotonously. Compared with associative
polymers, it is found that the magnitude of the transition
broadness is determined by the ratio of hydrophobic to hydrophilic
blocks, instead of chain length. When $\bar{\phi}_{\mathrm{P}}$ is
decreased, given a large $N_\textrm{st}$, the temperature-dependent range
of micellar aggregation and half-width of specific heat peak
initially decease, and then remain nearly constant. It is shown that
the transition broadness is concerned with the changes of the
relative magnitudes of the eductions of nonstickers and solvents
from micellar cores.

\pacs 61.25.Hp, 64.75.+g, 82.60.Fa
\end{abstract}

\section{Introduction}

Amphiphilic block copolymers are particularly versatile
macromolecules because they allow for a rich variety of different
structures. Their length and the number of blocks of each species
can be tuned at will, from di- and triblock to multiblock
copolymers. Their architectures can be linear, branched or
star-like, the blocks may be distributed randomly or regularly.
Consequently, amphiphilic block copolymers have  a great deal of
applications such as drug delivery vectors~\cite{Kata2001},
nanoparticle stabilizers, nanoreservoirs, emulsion stabilizers,
wetting agents, rheology modifiers~\cite{Ries2003,Bhat2001} or as
injectable scaffold materials for tissue engineering~\cite{Fati2008}.

Amphiphilic copolymers are capable of self-assembling into micelles when
temperature drops to a critical micelle temperature. Below the
critical micelle temperature there is an equilibrium region of a
certain width, where significant amounts of both free and associated
copolymer molecules coexist. Above the transition region most
copolymer molecules are in micelles. It is verified
theoretically~\cite{Han2012} and experimentally~\cite{Gold1997} that
the broad nature should be ascribed to the structural changes which
accompany the replacement of micellar core solvent by polymer.
However, the effect of the hydrophobic block on the structural
changes is not clarified so far, which is very important in high
polymer concentration regimes. The study of the effects of polymer
concentration and length of hydrophobic end block on the transition
broadness in amphiphilic ABA symmetric triblock copolymer solutions
would be quite useful to understand the thermodynamics of block copolymers
in a selective solvent.

The self-consistent field theory (SCFT) has been brought into use as a powerful tool in predicting the
morphologies of complex block
copolymers~\cite{Orland1996,Mats1994,Tang2004,He2004}. Recently, SCFT
has been applied to  study the properties of micelles in polymer
solutions~\cite{Cava2006,Jeli2007,Char2008}. In this report, a SCFT
lattice model is applied~\cite{Chen2006,Chen2007,Chen2008}. In
earlier publications, we have used the SCFT lattice model to study
the phase behavior of physically associating polymer
solutions~\cite{Han2010,Han2011,Han2012}. It is established that the
temperature-dependent behavior of aggregates is affinitive to chain
architecture~\cite{Han2012}, and the effect of polymer concentration is
in a way similar to that of chain architecture~\cite{Han2011}. Now
amphiphilic ABA symmetric triblock copolymer solutions are studied
using SCFT lattice model. The focus is made on the effects of the length of hydrophobic end
block and polymer concentration on the broadness of the transition
observed. It is found that the magnitude of the
transition broadness is related to the relative changes of the
eductions of nonsticker and solvent from micellar cores.

\section{Theory\label{sec2}}
This section briefly describes the self-consistent field theory
(SCFT) lattice model for $n_{\mathrm{P}}$ amphiphilic ABA symmetric triblock
copolymers which is assumed to be incompressible. Each triblock
molecule is composed of 2$N_\textrm{st}$ sticker segments forming two
hydrophobic end A block and $N_\textrm{ns}$ nonsticker segments forming the
hydrophilic middle B block, distributed over a lattice; the degree
of polymerization of the chain is $N$ ($=2N_\textrm{st}+N_\textrm{ns}$) and the total
number of lattice sites
is $N_{\mathrm{L}}$. In addition to polymer monomers, $%
n_{\mathrm{h}}$ solvent molecules are placed on the vacant lattice sites.
Stickers, nonsticky monomers and solvent molecules have the same
size and each occupies one lattice site, and thus $N_{\mathrm{L}}$ $=$ $%
n_{\mathrm{h}}+n_{\mathrm{P}}N$. Nearest neighbor pairs of stickers have attractive
interaction $-\epsilon$ with $\epsilon>0$, which is the only
non-bound interaction in the present system. The approximation of
the attractive interaction energy~\cite{Han2010} is expressed as:%
\begin{equation}
\frac{U}{k_\textrm{B}T}=-\chi \sum_{{r}}\widehat{\phi }_\textrm{st}({r})\widehat{\phi }%
_\textrm{st}({r}),  \label{01}
\end{equation}%
where $\chi $ is the Flory-Huggins interaction parameter in the
solutions, which equals $({z}/{2k_\textrm{B}T})\epsilon$, $z$ is the
coordination number of the lattice used, where $\sum_{r}$ means the
summation over all the lattice sites ${r}$ and. $\widehat{\phi
}_\textrm{st}({r})=\sum_{{j}}\sum_{s{\in \textrm{st}}}\delta
_{{r},{r}_{j,s}}$ is the volume fraction of stickers on the site ${r}$, where $j$ and $%
s$ are the indexes of chain and monomer of a polymer, respectively.
$s\in \textrm{st}$ means that the $s$th monomer belongs to the sticker
monomer type. In this simulation, we perform the SCFT calculations in the
canonical ensemble, and the field-theoretic free energy
$F$~\cite{Han2010,Fredr2005} is defined as
\begin{equation}
\frac{F[\omega _{+},\omega _{-}]}{k_\textrm{B}T}=\sum_{r}\left\{ \frac{1}{4\chi }%
\omega _{-}^{2}(r)-\omega _{+}(r)\right\} -n_{\mathrm{P}}\ln Q_{\mathrm{P}}[\omega
_\textrm{st},\omega _\textrm{ns}]-n_{\mathrm{h}}\ln Q_{\mathrm{h}}[\omega _{\mathrm{h}}],  \label{free0}
\end{equation}
where $Q_{\mathrm{h}}$ is the partition function of a solvent molecule
subject to the field $\omega _{\mathrm{h}}(r)=$ $\omega _{+}(r)$, which
is defined as $Q_{\mathrm{h}}=({1}/{n_{\mathrm{h}}})\sum_{r}\exp\left[ -\ \omega
_{\mathrm{h}}(r)\right]$. $Q_{\mathrm{P}}$ is the partition function of a
noninteraction polymer chain subject to the fields $\omega
_\textrm{st}(r)=\omega _{+}(r)-\omega _{-}(r)$ and $\omega
_\textrm{ns}(r)=\omega _{+}(r)$, which act on sticker and nonsticker
segments, respectively.

 Minimizing the free energy function $F$ with
$\omega _{-}(r)$ and $\omega _{+}(r)$ leads to the following
saddle point equations:
\begin{equation}
\omega _{-}(r)=2\chi \phi _\textrm{st}(r),  \label{scf1}
\end{equation}%
\begin{equation}
\phi _\textrm{st}(r)+\phi _\textrm{ns}(r)+\phi _{\mathrm{h}}(r)=1,
\end{equation}
where
\begin{equation}
\phi _\textrm{st}(r)=\frac{1}{zN_{\mathrm{L}}}\frac{n_{\mathrm{P}}}{Q_{\mathrm{P}}}\sum_{s\in \textrm{st}}%
\sum_{\alpha _{s}}\frac{G^{\alpha _{s}}(r,s|1)G^{\alpha _{s}}(r,s|N)}{G(r,s)%
}
\end{equation}
and
\begin{equation}
\phi _\textrm{ns}(r)=\frac{1}{zN_{\mathrm{L}}}\frac{n_{\mathrm{P}}}{Q_{\mathrm{P}}}\sum_{s\in \textrm{ns}}%
\sum_{\alpha _{s}}\frac{G^{\alpha _{s}}(r,s|1)G^{\alpha _{s}}(r,s|N)}{G(r,s)%
}
\end{equation}
are the average numbers of sticker and nonsticker segments at $r$,
respectively, and
\[
\phi_{\mathrm{h}}(r)=\frac{1}{N_{\mathrm{L}}}\frac{n_{\mathrm{h}}}{Q_{\mathrm{h}}}\exp \left[ - \omega _{\mathrm{h}}(r)\right]
\]
is the average
number of solvent molecules at $r$. $Q_{\mathrm{P}}$ is expressed as
\[
Q_{\mathrm{P}}=\frac{1%
}{zN_{\mathrm{L}}}\sum_{r_{N}}\sum_{{\alpha }_{N}}G^{\alpha _{N}}(r,N|1),
\]
where $r_{N}$ and ${\alpha }_{N}$ denote the position and
orientation of the $N$th segment of the chain, respectively.
$\sum_{r_{N}}\sum_{\alpha _{N}}$ means the summation over all the
possible positions and orientations of the $N$th segment of the
chain, respectively. $G^{\alpha _{s}}(r,s|1)$ and $G^{\alpha _{s}}(r,s|N)$ are the
end segment distribution functions of the $s$th segment of the
chain. $G(r,s)$ is the free segment weighting factor.

Following the scheme of Schentiens and Leermakers~\cite{Leer1988},
$G^{\alpha _{s}}(r,s|1)$ is the end segment distribution function of
the $s$th segment of the chain, which is evaluated from the
following recursive relation:
\begin{equation} G^{\alpha
_{s}}(r,s|1)=G(r,s)\sum_{r_{s-1}^{\prime }}\sum_{\alpha
_{s-1}}\lambda _{r_{s}-r_{s-1}^{\prime }}^{\alpha _{s}-\alpha
_{s-1}}G^{\alpha _{s-1}}(r^{\prime },s-1|1), \label{free}
\end{equation}
where $G(r,s)$ is the free segment
weighting factor and is expressed as
\[G(r,s)=\left\{
\begin{array}{ll}
\exp\left[-\omega_\textrm{ns}(r_{s})\right], &  s\in \textrm{ns},\\
\exp\left[-\omega _\textrm{st}(r_{s})\right], & s\in \textrm{st}.
\end{array}
\right.
\]
The initial condition is $G^{\alpha _{1}}(r,1|1)=G(r,1)$ for
all the values of $\alpha _{1}$. In the above expression, the values
of $\lambda_{r_{s}-r'_{s-1}}^{\alpha _{s}-\alpha _{s-1}}$ depend
on the chain model used. We assume that
\[
\lambda_{r_{s}-r^{'}_{s-1}}^{\alpha _{s}-\alpha _{s-1}}=
\left\{
\begin{array}{ll}
0, & \alpha _{s}=\alpha _{s-1}\,,\\
{1}/(z-1), & \text{otherwise}\,.
\end{array}
\right.
\]
Another end segment distribution function $G^{\alpha _{s}}(r,s|N)$ is
evaluated from the following recursive relation:
\begin{equation}
G^{\alpha _{s}}(r,s|N)=G(r,s)\sum_{r_{s+1}^{\prime }}\sum_{\alpha
_{s+1}}\lambda _{r_{s+1}^{\prime }-r_{s}}^{\alpha _{s+1}-\alpha
_{s}}G^{\alpha _{s+1}}(r^{\prime },s+1|N),
\end{equation}
with the initial condition $G^{\alpha _{N}}(r,N|N)=G(r,N)$ for all
the values of $\alpha _{N}$. In this work, the chain is described
as a random walk without the possibility of direct backfolding.
Although self-intersections of a chain are not permitted, the
excluded volume effect is sufficiently taken into
account~\cite{Medv2001}.

The saddle point is calculated using the pseudo-dynamical evolution
process~\cite{Han2010}.
The calculation is initiated from appropriately random-chosen fields $%
\omega_{+}(r)$ and $\omega _{-}(r)$, and stopped when the change
of free energy $F$ between two successive iterations is reduced to
the needed precision. The resulting configuration is taken as a
saddle point one. By comparing the free energies of the saddle point
configurations obtained from different initial fields, the relative
stability of the observed morphologies can be assessed.

\section{Result and discussion\label{sec3}}

In our studies, the amphiphilic ABA symmetric triblock copolymers
depend on three tunable molecular parameters: $\chi$ (the
Flory-Huggins interaction parameter), $N_\textrm{ns}$ (the length of
hydrophilic middle block, in this paper $N_\textrm{ns}=9$ ) and $N_\textrm{st}$
(the length of each hydrophobic end block). The simulation
calculations are performed in a three-dimensional simple cubic lattice
with periodic boundary condition. The results presented below are
obtained from the lattice with $N_{\mathrm{L}}=26^{3}$. The focus is made on the temperature
behavior of micelle morphologies when the length of
hydrophobic end block changes.

\begin{figure}[!t]
\centerline{
\includegraphics[width=0.5\textwidth]{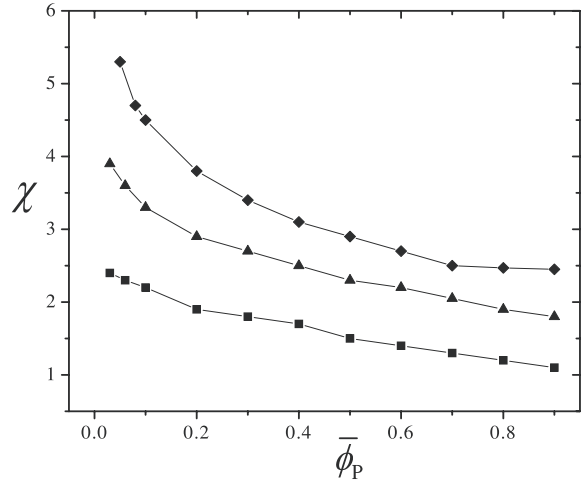}
}
\caption{The
phase diagram for the systems with different lengths of each
hydrophobic end block $N_\textrm{st}$. The boundary between homogenous
solutions (blow boundary) and micelle morphology (above boundary) is
obtained. The squares, triangles and diamonds correspond to the
boundaries for $N_\textrm{st}=4, \ 2, \ 1$, respectively.\label{phadia}}
\end{figure}

Figure~\ref{phadia} shows the phase diagram of the systems with
different length of each hydrophobic end block $N_\textrm{st}$. When $\chi$
is increased, micelles are observed as a inhomogenous morphology
if $N_\textrm{st}=1$\,\footnote{The structural morphology of MFH morphology~\cite{Han2010} occurs at a
narrow region of $\Delta\chi=0.1 $ neighboring the micellar boundary when $\bar{\phi}_{\mathrm{P}}>0.7$ (not
shown), which is ignored.}. The $\chi$ value on micellar boundary
increases with decreasing $\bar{\phi}_{\mathrm{P}}$. When $N_\textrm{st}$ is
increased, at fixed $\bar{\phi}_{\mathrm{P}}$, the $\chi$ value on micellar
boundary shifts to a small value. The increase in the length of
hydrophobic end block is favorable to the occurrence of micelles in
the system.

\begin{figure}[!b]
\centerline{
\includegraphics[width=0.48\textwidth]{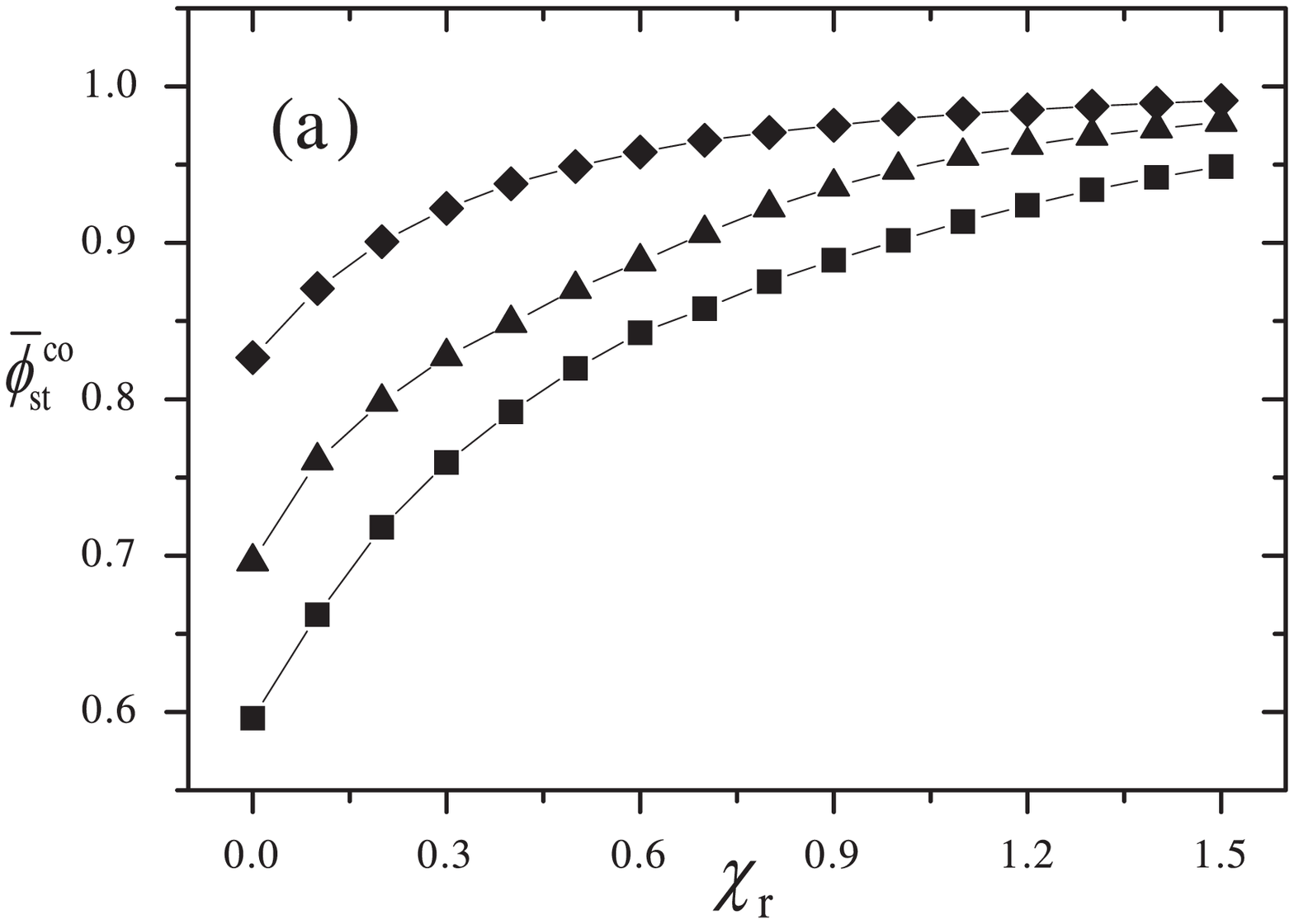}
\hspace{1mm}
\includegraphics[width=0.48\textwidth]{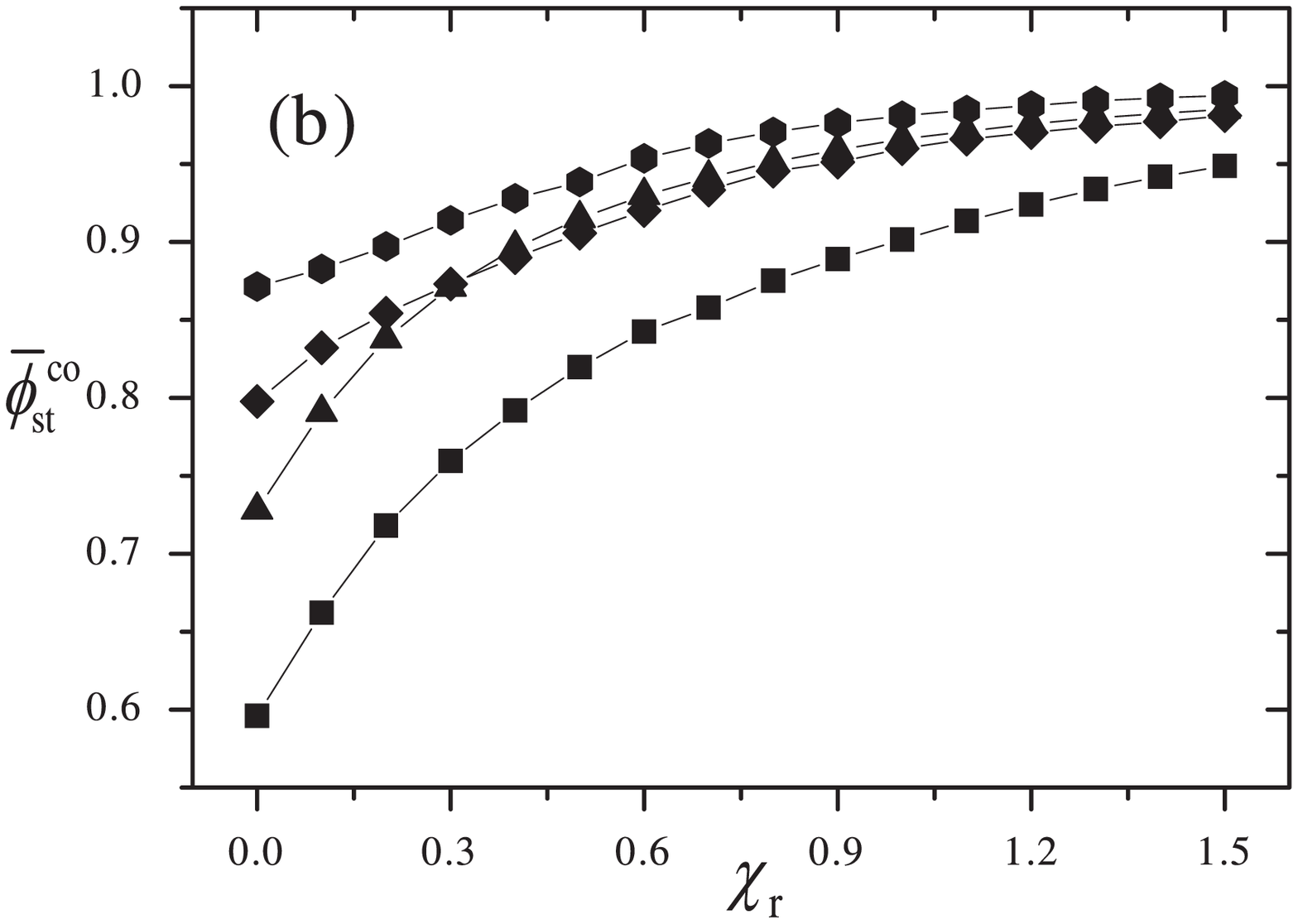}
}
\caption{The variations of average numbers of stickers at the
micellar cores with the $\chi$ deviation from micellar boundary
$\chi_r$ , for various lengths of each hydrophobic end block
$N_\textrm{st}$ at $\bar{\phi}_{\mathrm{P}}=0.8$ and different $\bar{\phi}_{\mathrm{P}}$ at
$N_\textrm{st}=4$, are presented by figure~\ref{concen}~(a) and (b),
respectively. In figure~(a), The squares, triangles and diamonds
correspond to $N_\textrm{st}=4,\ 2,\ 1$, respectively; In figure~(b), the
squares, triangles, diamonds and hexagons denote
$\bar{\phi}_{\mathrm{P}}=0.8,\ 0.4,\ 0.2,\ 0.1$, respectively. \label{concen}}
\end{figure}

The variation of the average number of stickers at the micellar cores
denoted by $\bar{\phi} _\textrm{st}^\textrm{co}$ with the $\chi$ deviation from
micellar boundary $\chi_r$ is calculated. For different $N_\textrm{st}$ at
$\bar{\phi}_{\mathrm{P}}=0.8$ and different $\bar{\phi}_{\mathrm{P}}$ at
$N_\textrm{st}=4$, the curves of $\bar{\phi} _\textrm{st}^\textrm{co}(\chi_r)$  are shown
in figure~\ref{concen}~(a) and (b), respectively. When $N_\textrm{st}=1$,
as shown in figure~\ref{concen}~(a), $\bar{\phi} _\textrm{st}^\textrm{co}$ rises and
approaches to 1 with an increase in $\chi_r$. When $N_\textrm{st}$ is
increased, the value of $\bar{\phi} _\textrm{st}^\textrm{co}$ at fixed $\chi_r$
decreases, and its temperature-dependent range goes up. For
$N_\textrm{st}=4$ [see figure~\ref{concen}~(b)], when $\bar{\phi}_{\mathrm{P}}$
is decreased, the value of $\bar{\phi} _\textrm{st}^\textrm{co}$ at fixed $\chi_r$
does monotonously increase only in the $\chi_r$ range near
$\chi_r=0$. When $\chi_r$ is increased to some extent, at the middle
concentration regimes, $\bar{\phi} _\textrm{st}^\textrm{co}$ at fixed $\chi_r$
decreases with a decreasing $\bar{\phi}_{\mathrm{P}}$. It is shown that the
magnitude of the temperature-dependent range of micelle aggregation
does not monotonously change with $\bar{\phi}_{\mathrm{P}}$, which is
different from the case of changing $N_\textrm{st}$ at fixed
$\bar{\phi}_{\mathrm{P}}$.

\begin{figure}[!t]
\centerline{
\includegraphics[width=0.48\textwidth]{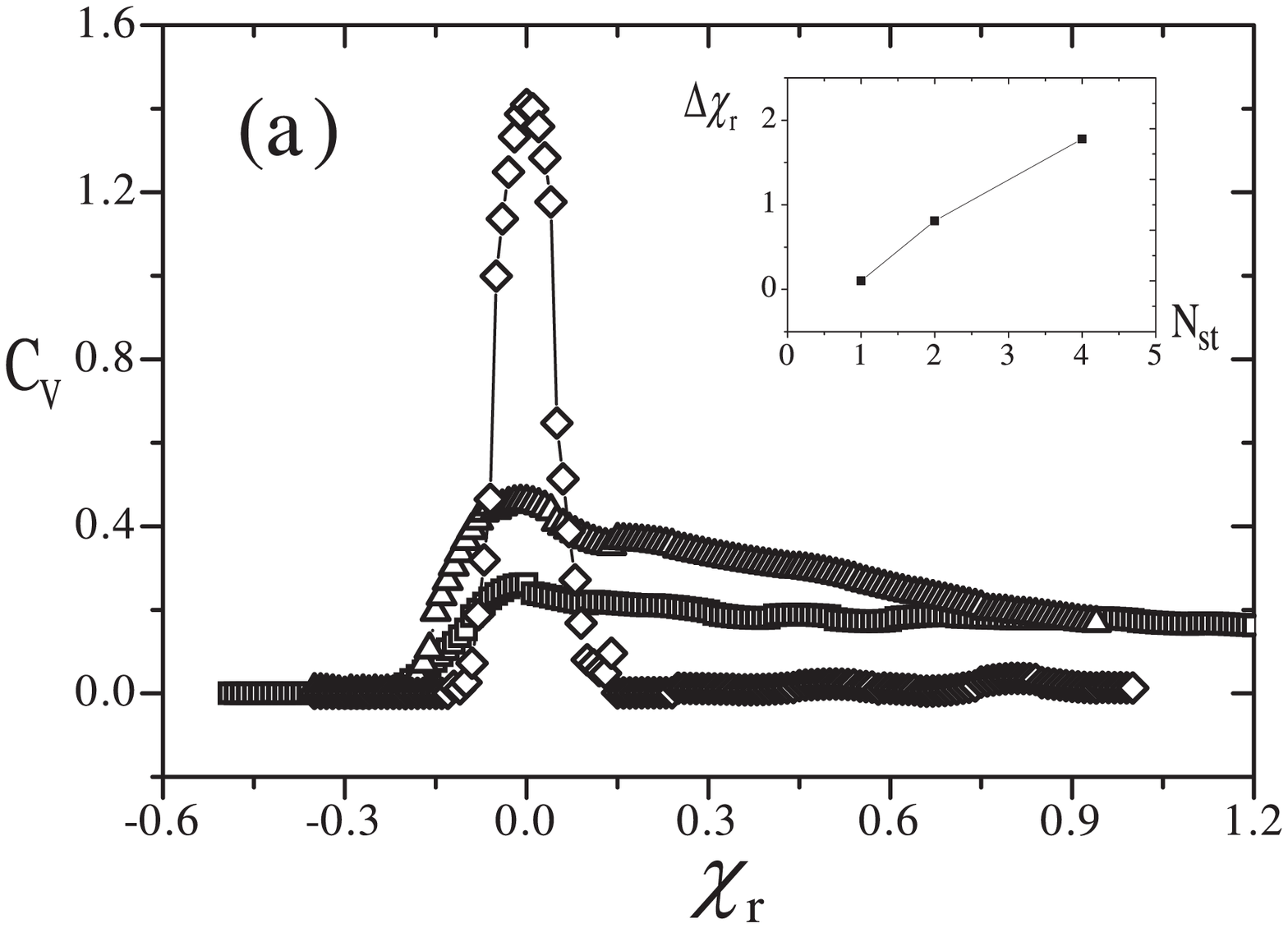}
\hspace{1mm}
\includegraphics[width=0.48\textwidth]{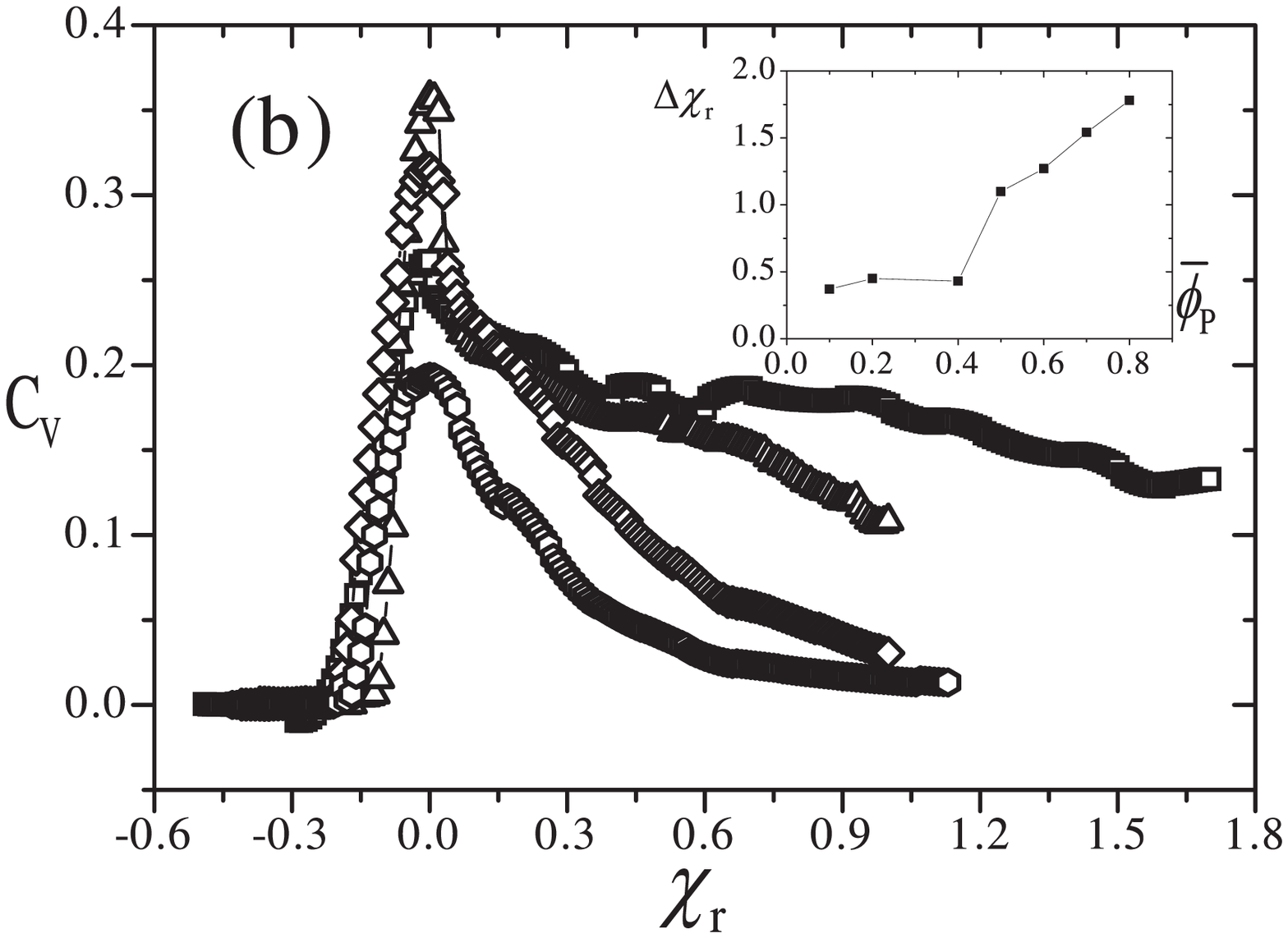}
}
\caption{The changes of specific heat capacity in different
amphiphilic ABA symmetric triblock copolymers with the $\chi$
deviation from micellar boundary $\chi_r$ are presented by figure~\ref{Cv}~(a) and (b) corresponding to the systems shown by figure~\ref{concen}~(a) and (b), respectively.  \label{Cv}}
\end{figure}

The half-width of a specific
heat peak may be an intrinsic measure of transition
broadness.~\cite{Doug2006,Han2012}. In this work, the heat capacity
 per site of amphiphilic ABA symmetric triblock copolymers is
expressed as (in the unit of $k_{\textrm{B}}$):
\begin{eqnarray} \label{scf5-2}
C_{V} =\left(\frac{\partial {U}}{\partial
{T}}\right)_{N_{\mathrm{L}},{n}_{\mathrm{P}}}
=\frac{1}{N_{\mathrm{L}}}\chi ^{2}\frac{\partial }{\partial {\chi
}}\left[ \sum_{r}\phi _\textrm{st}^{2}(r)\right] .
\end{eqnarray}%
For various $N_\textrm{st}$ at $\bar{\phi}_{\mathrm{P}}=0.8$ and different
concentrations at $N_\textrm{st}=4$, the $C_{V}(\chi_r)$ curves of the
unimer-micelle transition  are shown in figure~\ref{Cv}~(a) and
\ref{Cv}~(b), respectively. For unimer-micelle transition, a peak
appears in each $C_{V}(\chi_r )$ curve. When $N_\textrm{st}$ is increased,
as shown in figure~\ref{Cv}~(a), the half-width of the transition peak
rises, and the symmetry and the height of the transition peak
decrease. The broadness of unimer-micelle transition increases with
increasing the length of hydrophobic end block, which is in
reasonable agreement with that on temperature-dependent range of
micellar aggregation. Whereas for the case of changing
$\bar{\phi}_{\mathrm{P}}$ at $N_\textrm{st}=4$, the half-width and height of the
transition peak do not monotonously change with $\bar{\phi}_{\mathrm{P}}$.
When $\bar{\phi}_{\mathrm{P}}$ is decreased, the symmetry of the peak
always increases, the
 height of the transition peak firstly increases, and then
 decreases. Its half-width initially drops, and
then nearly remains constant with a decreasing $\bar{\phi}_{\mathrm{P}}$. In
other words, at high concentrations the broadness of unimer-micelle
transition is affected by polymer concentration. In middle and low
concentration regimes, the height of the transition peak is affected
by the change of polymer concentration. However, the transition
broadness is almost unrelated to polymer concentration. It is
shown that the effect of polymer concentration on the transition
broadness is consistent with that on the temperature-dependent range
of micellar aggregation.

 It is obvious that  the increase in the degree of aggregation at
micellar cores results from the eductions of nonsticky monomers and
solvents. Micelles appear when temperature drops below the critical micelle temperature. With a further decrease of temperature, solvents and
nonstickers continue to be expelled from micellar cores, and the
degree of aggregation of micellar cores strengthens. Therefore, the
temperature-dependent behavior of micellar aggregation brings about
the existence of the transition broadness, rather than a transition
point. The broadness of unimer-micelle transition increases with
increasing the length of hydrophobic end block (i.e., the length of
chain), which is consistent with the effect of decreasing the length
of hydrophilic middle block between neighboring hydrophobic blocks,
at a fixed chain length, in associative polymer
solutions~\cite{Han2012}. It is demonstrated that the broadness of
the transitions concerned with micelles is determined by  the ratio
of hydrophobic to hydrophilic blocks, which is not related to the
length of polymer chain.

\begin{figure}[!h]
\centerline{
\includegraphics[width=0.6\textwidth]{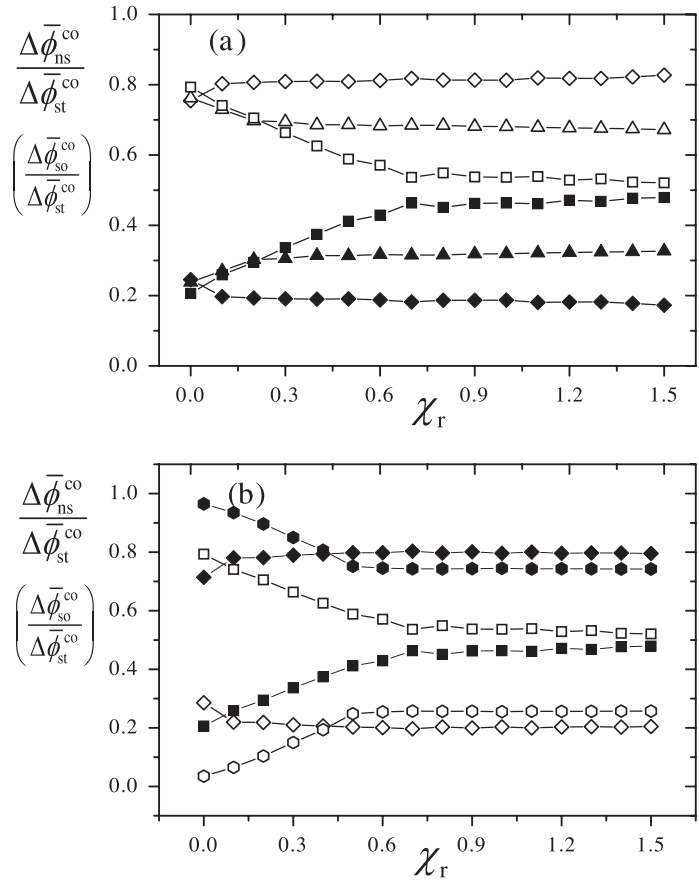}
}
\caption{The variation of the ratios of the changes of average
numbers of nonstickers and solvents to that of stickers at the
micellar cores when the $\chi$ deviation from micellar boundary
$\chi_r$, for various lengths of each hydrophobic end block $N_\textrm{st}$
at $\bar{\phi}_{\mathrm{P}}=0.8$ and different $\bar{\phi}_{\mathrm{P}}$ at
$N_\textrm{st}=4$, is presented by figure~\ref{percent}~(a) and (b),
respectively. The change $\Delta\phi^\textrm{co}_{s}(\chi_r)$ equals
$\phi^\textrm{co}_{s}(\chi_r)-\phi^\textrm{co}_{s}(\chi_r-0.1)$, where $s$ denotes
$\mathrm{st}$, $\mathrm{ns}$, $\mathrm{so}$, respectively. In figure~(a), The open and solid
squares, triangles and diamonds correspond to the cases for
$N_\textrm{st}=4, 2, 1$, respectively; In figure (b), the open and solid
squares, diamonds
 and hexagons denote  the cases for
$\bar{\phi}_{\mathrm{P}}=0.8,\ 0.4,\ 0.1$, respectively. \label{percent}}
\end{figure}
Furthermore, the relative magnitude of contributions of nonsticky
monomers and solvents to aggregation of micellar cores should be
related to micelle structure and the relationship among micelles. At
high concentrations, when the length of hydrophobic end block
$N_\textrm{st}$ is increased, the micellar volume fraction in the system
rises and the micelle structure tends to be intricate. These factors
result in the difficulties of the eductions of nonsticky monomers
and solvents from micellar cores. Therefore, with an increase in
$N_\textrm{st}$ at fixed $\bar{\phi}_{\mathrm{P}}$, the temperature-dependent
range of aggregation of micellar cores, as well as the transition
broadness, rises. Moreover, when $N_\textrm{st}$ is relatively big, the
relationship among micelles is strong, which markedly hampers the
eduction of nostickers, thus the contribution of solvents will be
rather important. As shown in figure~\ref{percent}~(a), when
$N_\textrm{st}$ is increased from $N_\textrm{st}=1$, given a fixed
$\bar{\phi}_{\mathrm{P}}$, the contribution of nonstickers to aggregation
of stickers goes down and that of solvents rises with an increasing
$\chi_r$, at the neighborhood of $\chi_r>0$. It is noted that the
evidently temperature-dependent range of the ratios of the changes
of average numbers of nonstickers and solvents to that of stickers
at the micellar cores rises with an increasing $N_\textrm{st}$. The larger is
the evidently temperature-dependent range of the above ratio, the
bigger is the transition broadness. It is shown that the magnitude
of the transition broadness is concerned with the changes of the
relative magnitudes of the eductions of nonstickers and solvents
from micellar cores.

In high concentrations, when polymer concentration is decreased, for
a large $N_\textrm{st}$, the effect of the relationship among micelles on
eductions of nonstickers and solvents evidently dies down.
Therefore, the transition broadness decreases with a decreasing
$\bar{\phi}_{\mathrm{P}}$. At intermediate and low concentrations, the
effect of the relationship among micelles on eductions of
nonstickers and solvents is weak, especially to nonstickers. Seen
from figure~\ref{percent}~(b), with an increasing $\chi_r$ from
$\chi_r=0.1$, the ratios of the changes of average numbers of
nonstickers and solvents to that of stickers at the micellar cores
nearly remain constant, where solvents and nonstickers are expelled
proportionally. When polymer concentration is decreased to some
extent, the aggregation of micellar cores is dominated by the
eductions of solvents. Due to the existence of a large quantity of
solvents among micellar cores, it is difficult to expel a small amount of
solvent at the micellar core. Therefore, although the
effect of the relationship among micelles is already very weak, the
transition broadness always remains constant with a decrease in
$\bar{\phi}_{\mathrm{P}}$ when the contribution of nonstickers
is rather important and is temperature-dependent in a larger range of
$\chi_r$ [figure~\ref{percent}~(b)].

\section{Conclusion and summary\label{sec4}}
The effects of the length of  each hydrophobic end block $N_\textrm{st}$
and polymer concentration $\bar{\phi}_{\mathrm{P}}$ on the transition
broadness in amphiphilic ABA symmetric triblock copolymer solutions
are studied using the self-consistent field lattice model. When
$N_\textrm{st}$ is changed, at fixed  $\bar{\phi}_{\mathrm{P}}$, micelles occur
at a higher temperature, and the broadness of unimer-micelle
transition also increases. Compared with associating polymer
solutions, it is found that the magnitude of the transition
broadness is determined by the ratio of hydrophobic to hydrophilic
blocks rather than by the length of polymer chain. When
$\bar{\phi}_{\mathrm{P}}$ is decreased, given a large $N_\textrm{st}$, the
transition does not change monotonously. In high concentration
regimes, the transition broadness decreases with decreasing
$\bar{\phi}_{\mathrm{P}}$, and in intermediate and low concentration
regimes, the transition broadness remains constant with
$\bar{\phi}_{\mathrm{P}}$. It is demonstrated that the magnitude of the
transition broadness  is concerned with the changes of the relative
magnitudes of the eductions of nonstickers and solvents from
micellar cores.

\section*{Acknowledgements}
This research is financially supported
by the National Nature Science Foundations of China (11147132) and
the Inner Mongolia municipality (2012MS0112), and the Innovative
Foundation of Inner Mongolia University of Science and Technology
(2011NCL018).

\ukrainianpart

\title{Вплив концентрації полімера і довжини гідрофобного  прикінцевого
блоку на ширину переходу мономер-міцела в  ABA симетричних триблочних
амфіфільних кополімерних розчинах}

\author{К.-Г. Ган\refaddr{label1,label2}, Й.-Г. Ма\refaddr{label1,label2}, С.-Л. Оуянг\refaddr{label2}}
\addresses{
\addr{label1} Школа математики, фізики і біологічної інженерії,
університет науки і технологій внутрішньої Монголії, Баоту 014010,
Китай
\addr{label2} Головна лабораторія інтегрованого використання
мультиметалічних ресурсів Баян Обо, університет науки і технології
внутрішньої Монголії, Баоту 014010, Китай }

\makeukrtitle

\begin{abstract}
\tolerance=3000%
Вплив довжини кожного гідрофобного  прикінцевого блоку
$N_\textrm{st}$ і концентрації полімера $\bar{\phi}_{\mathrm{P}}$ на ширину
переходу в симетричних  ABA триблочних амфіфільних кополімерних
розчинах досліджується шляхом використання ґраткової моделі
самоузгодженого поля. Коли система охолоджена, спостерігаються
міцели, тобто відбувається перехід однорідний розчин
(мономер)-міцела. Якщо $N_\textrm{st}$ зростає при сталому
$\bar{\phi}_{\mathrm{P}}$, то міцели виникають при високій температурі, а
температурно залежна область агрегації міцел і півширина піку
питомої теплоємності для переходу мономер-міцела зростають
монотонно. Порівнюючи з асоціативними полімерами, знайдено, що
величина ширини переходу  визначається відношенням гідрофобних блоків до
гідрофільних, а не довжиною ланцюга. Коли $\bar{\phi}_{\mathrm{P}}$
зменшується при великому значенні $N_\textrm{st}$, температурно
залежна область міцелярної агрегації та півширина піку питомої
теплоємності спочатку зменшуються, а потім залишаються майже
сталими. Показано, що ширина переходу пов'язана зі зміною
відносних величин виділення незв'язувальної речовини і розчинників з
міцелярних корів.
\keywords ширина переходу, самоузгоджене поле, амфіфільний кополімер
\end{abstract}


\begin{thebibliography}{99}

\bibitem{Kata2001} Kataoka K., Harada A., Nagasaki Y., Adv. Drug Delivery Rev.,
2001, \textbf{47}, 113; \doi{10.1016/S0169-409X(00)00124-1}.

\bibitem{Ries2003} Riess G., Prog. Polym. Sci., 2003, \textbf{28}, 1107; \doi{10.1016/S0079-6700(03)00015-7}.

\bibitem{Bhat2001} Bhatia S.R., Mourchid A., Joanicot M., Curr. Opin. Colloid
Interface Sci., 2001, \textbf{6}, 471; \\ \doi{10.1016/S1359-0294(01)00122-4}.

\bibitem{Fati2008} Fatimi A., Tassin J.F., Quillard S. ,  Axelos M.A.V.,  Weiss P.,
Biomaterials, 2008, \textbf{29}, 533; \\ \doi{10.1016/j.biomaterials.2007.10.032}.

\bibitem{Han2012} Han X.G., Zhang X.F., Ma Y.H., Condens. Matter Phys.,
2012, \textbf{15}, No.~3, 33602; \doi{10.5488/CMP.15.33602}.

\bibitem{Gold1997}Goldmints I., Gottberg F.K.V., Smith K.A., Hatton T.A.,
Langmuir, 1997, \textbf{13}, 3659; \doi{10.1021/la970140v}.


\bibitem{Orland1996} Orland H., Schick M., Macromolecules, 1996, \textbf{29}, No.~2, 713; \doi{10.1021/ma9508461}.

\bibitem{Mats1994} Matsen M.W., Schick M., Phys. Rev. Lett., 1994, \textbf{72}, No.~16, 2660; \doi{10.1103/PhysRevLett.72.2660}.

\bibitem{Tang2004} Tang P., Qiu F., Zhang H.D., Yang Y.L., Phys. Rev. E, 2004, \textbf{69}, No.~3, 031803.

\bibitem{He2004} He X., Liang H., Huang L., Pan C., J. Phys. Chem. B, 2004, \textbf{108}, No.~5, 1731; \doi{10.1021/jp0359337}.


\bibitem{Cava2006} Cavallo A., Muller M., Binder K., Macromolecules, 2006, \textbf{39}, No.~26, 9539; \doi{10.1021/ma061493g}.

\bibitem{Jeli2007} Jelinek K., Limpouchova Z., Uhlk F., Prochazka K., Macromolecules, 2007, \textbf{40}, No.~21, 7656; \\ \doi{10.1021/ma070928c}.

\bibitem{Char2008} Charlaganov M., Borisov O.V., Leermakers F.A.M., Macromolecules, 2008, \textbf{41}, 3668; \doi{10.1021/ma800130q}.

\bibitem{Chen2006} {Chen J.Z., Zhang C.X., Sun Z.Y., Zheng Y.S.,
 An L.J., J. Chem. Phys., 2006, \textbf{124}, 104907; \doi{10.1063/1.2176619}.}

\bibitem{Chen2007} Chen J.Z., Sun Z.Y., Zhang C.X., An L.J., Tong Z.,
 J. Chem. Phys., 2007, \textbf{127}, 024105; \doi{10.1063/1.2750337}.

\bibitem{Chen2008} Chen J.Z., Sun Z.Y., Zhang C.X., An L.J., Tong Z., J. Chem. Phys., 2008, \textbf{128}, 074904; \doi{10.1063/1.2831802}.

\bibitem{Han2010} Han X.G., Zhang C.X., J. Chem. Phys., 2010, \textbf{132}, 164905; \doi{10.1063/1.3400648}.

\bibitem{Han2011} Han X.G., Zhang X.F., Ma Y.H., Zhang C.X.,
 Guan Y.B., Condens. Matter Phys., 2011, \textbf{14}, No.~4, 43601; \\ \doi{10.5488/CMP.14.43601}.

\bibitem{Fredr2005}  Fredrickson G.H., The Equilibrium Theory of Inhomogenous Polymers,
Clarendon Press, Oxford, 2005.

\bibitem{Leer1988} Leermakers F.A.M., Scheutjens J.M.H.M., J. Chem. Phys., 1988, \textbf{89}, No.~5, 3264; \doi{10.1063/1.454931}.

\bibitem{Medv2001} Medvedevskikh Y.G., Condens. Matter Phys., 2001, \textbf{4}, No.~2, 209; \doi{10.5488/CMP.4.2.209}.


\bibitem{Doug2006} Douglas J.F., Dudowicz J., Freeda K.F., J. Chem. Phys., 2006, \textbf{125}, 114907; \doi{10.1063/1.2356863}.

\end{thebibliography}
\end{document}